\documentclass{iopart}
\usepackage{epsfig}
\usepackage{bm}

\def\Journal#1#2#3#4#5{#1 {\it #2} #3 {\bf #4} #5}
\def\Preprint#1#2{#1 {\it Preprint} #2}

\def\NP{Nucl. Phys.}
\def\PL{Phys. Lett.}
\def\PR{Phys. Rev.}
\def\PRL{Phys. Rev. Lett.}

\newcommand{\eq}{{\,=\,}}

\newcommand{\gapp}{\,{\raisebox{-.2ex}{$\stackrel{>}{_\sim}$}}\,}
\newcommand{\lapp}{\,{\raisebox{-.2ex}{$\stackrel{<}{_\sim}$}}\,}

\begin{document}

\title[Hydrodynamics at RHIC]{Hydrodynamics at RHIC -- how well does 
it work, where and how does it break down?}

\author{Ulrich Heinz\dag\footnote[3]{email: heinz@mps.ohio-state.edu. 
 Work supported by the U.S. Department of Energy under Grant 
 No. DE-FG02-01ER41190} 
}
\address{\dag\ 
Physics Department, The Ohio State University, Columbus, OH 43210}

\begin{abstract}
I review the successes and limitations of the ideal fluid dynamic model
in describing hadron emission spectra from Au+Au collisions at the 
Relativistic Heavy Ion Collider (RHIC).
\end{abstract}

%%%%%%%%%%%%%%%%%%%%%%%%%%%%%%%%%%%%%%%%%%%%%%%%%%%%%%%%%%%%%%%%%%%%%%%%%%%%
\section{Introduction}
%%%%%%%%%%%%%%%%%%%%%%%%%%%%%%%%%%%%%%%%%%%%%%%%%%%%%%%%%%%%%%%%%%%%%%%%%%%%

For the last 6-8 months, an intense discussion has been going on in the 
relativistic heavy-ion collision (RHIC) community whether one can consider 
the discovery of the quark-gluon plasma (QGP) as settled and if it is 
time for the RHIC program to move on, entering a new phase of detailed 
and precise measurements to explore the quan\-ti\-ta\-tive properties
of this new state of hot and dense matter \cite{TheoryWhitePapers,white}. 
Whereas theorists seem to generally agree that QGP has been successfully 
created in Au+Au collisions at RHIC \cite{TheoryWhitePapers,Heinz:2001xi}, 
the experimental collaborations are more cautious in their assessments 
\cite{white}. General agreement exists, however, that RHIC has produced 
ther\-ma\-lized matter at unprecedented high energy density. This 
conclusion is based on the
successful hydrodynamic description of the measured hadron momentum 
spectra, in particular the elliptic flow and its ``fine structure'',
i.e. its dependence on the hadron mass \cite{Kolb:2003dz}.

Hydrodynamics provides a direct link between the equation of state 
(EOS) of the expanding fluid and the flow pattern manifested in the 
emitted hadron spectra. Just like any other phase transition, the
quark-hadron phase transition in QCD generates a ``soft'' region in
the EOS within which the ability of the pressure to generate flow is 
greatly reduced. This is expected to leave visible traces in the
{\em flow excitation function}. For example, hydrodynamics predicts
that the elliptic flow coefficient $v_2$ should show a non-monotonic 
dependence on the collision energy \cite{KSH00}, arising from a 
reduction in the elliptic flow signal if the collision fireball 
spends the crucial early stage, where such flow would be created, in 
the phase transition region.

A quantitative determination of the EOS requires precision flow data
as well as systematic theoretical studies of the influence of the 
initial conditions, equation of state, non-ideal transport effects, 
and the final 
decoupling kinetics on the observed hadron spectra. With the present
theory and RHIC data, we are just at the beginning of such a program.
That the RHIC data so far provide strong support for the basic validity 
of the hydrodynamic approach is a good omen, raising the hope that such
quantitative studies will be ultimately successful. There would be
much less optimism if RHIC had repeated the history of lower-energy
heavy-ion collisions where hydrodynamics never worked well enough
to provide more than a rough qualitative understanding of the 
collision dynamics. When the fireball dynamics proceeds too 
far away from local thermal equilibrium, it requires a microscopic 
transport theoretical description, involving a variety of
non-equilibrium mechanisms with typically poorly constrained parameters.
This renders the extraction of the equation of state difficult and 
uncertain. 

Of course, even at RHIC the hydrodynamic approach has its limitations.
Ideal fluid dynamics, which assumes instantaneous local
thermal equilibration, always remains an idealization; however, 
quantitative control of possibly small non-ideal effects due to shear
and bulk viscosity, heat conduction and diffusion requires a numerical
implementation of {\em viscous relativistic hydrodynamics} which has its
technical and conceptual difficulties and is only now beginning to
become practical \cite{Muronga}. Furthermore, in very peripheral 
heavy-ion collisions, where the size and lifetime of the collision 
fireball become small, even a viscous hydrodynamic approach is expected
to eventually break down. Also, particles with very large transverse
momenta (jets) are never expected to suffer sufficiently many 
interactions with the fireball medium to fully thermalize before 
escaping; hence a hydrodynamic approach can never work at very
high $p_\perp$. However, we can turn this inescapable failure of
hydrodynamics in small collision systems and at high $p_\perp$ to 
our favor: since ideal fluid dynamics appears to work well in 
near-central collisions, near mid-rapidity, and at low 
$p_\perp\lapp1.5-2$\,GeV/$c$ \cite{Kolb:2003dz}, we can study its 
gradual breakdown at larger impact parameters, rapidities, and 
transverse momenta in order to learn something about the mechanisms 
for the {\em approach to thermal equilibrium} at the beginning of 
the collision and the {\em decay of thermal equilibrium} near the 
end of the expansion stage, and hence about the transport properties   
of the early quark-gluon plasma and the late hadron resonance gas
created in these collisions. In this talk I will lay out the beginnings
of such a program.

%%%%%%%%%%%%%%%%%%%%%%%%%%%%%%%%%%%%%%%%%%%%%%%%%%%%%%%%%%%%%%%%%%%%%%%%%%%%
\section{Hydrodynamic description of single particle spectra and elliptic
flow}
%%%%%%%%%%%%%%%%%%%%%%%%%%%%%%%%%%%%%%%%%%%%%%%%%%%%%%%%%%%%%%%%%%%%%%%%%%%%

If the dense matter formed in the nuclear collision thermalizes locally
on a time scale much shorter than any macroscopic dynamical scale, its
evolution can be described by ideal fluid dynamics. The fluid's state is
then completely determined by the space-time profiles for its energy
density $e(x)$, baryon density $n(x)$, and flow 4-velocity $u^\mu(x)$.
The equation of state $p(e,n)$ closes the set of equations. Small 
deviations from local thermal 
equilibrium result in viscous corrections to the ideal fluid decomposition 
of energy-momentum tensor and baryon current, involving gradients of the 
above profiles multiplied by transport coefficients. For large deviations
from local thermal equilibrium this expansion in terms of gradients
of $e$, $n$ and $u^\mu$ breaks down, and the macroscopic hydrodynamic
approach must be replaced by a microscopic kinetic one which evolves
the distribution functions in both coordinate and momentum space.
 
Ideal fluid dynamics does not work during the initial collision stage,
when the two colliding nuclei penetrate each other, converting some
of their initial kinetic energy into creation of new matter, and it 
is also inapplicable during the last stage of the collision when
the expanding matter becomes so dilute that interactions among
its constituents are rare and local equilibrium breaks down. The 
hydrodynamic model therefore needs {\em initial conditions} (supposedly
the result of an early pre-equilibrium quantum kinetic evolution
leading to initial thermalization) and {\em final conditions} (a freeze-out
criterium describing the transition from a locally thermalized fluid
to an ensemble of free-streaming, noninteracting hadrons). Both initial
and final conditions involve free parameters which must be tuned to data.
As described in some detail in \cite{angra}, they can all be fixed by
only using data on the collision centrality dependence of the charged 
particle multiplicity and the {\em pion} and {\em proton} transverse 
momentum spectra in {\em central} ($b\eq0$) collisions. All other hadron 
spectra from central collisions, and all features of all hadron spectra 
in non-central collisions (in particular the elliptic flow for all 
hadron species) are then parameter-free predictions of the model 
\cite{HKHRV01}.

\begin{figure}[htb]
\begin{minipage}[t]{62mm}
 \includegraphics[width=62mm]{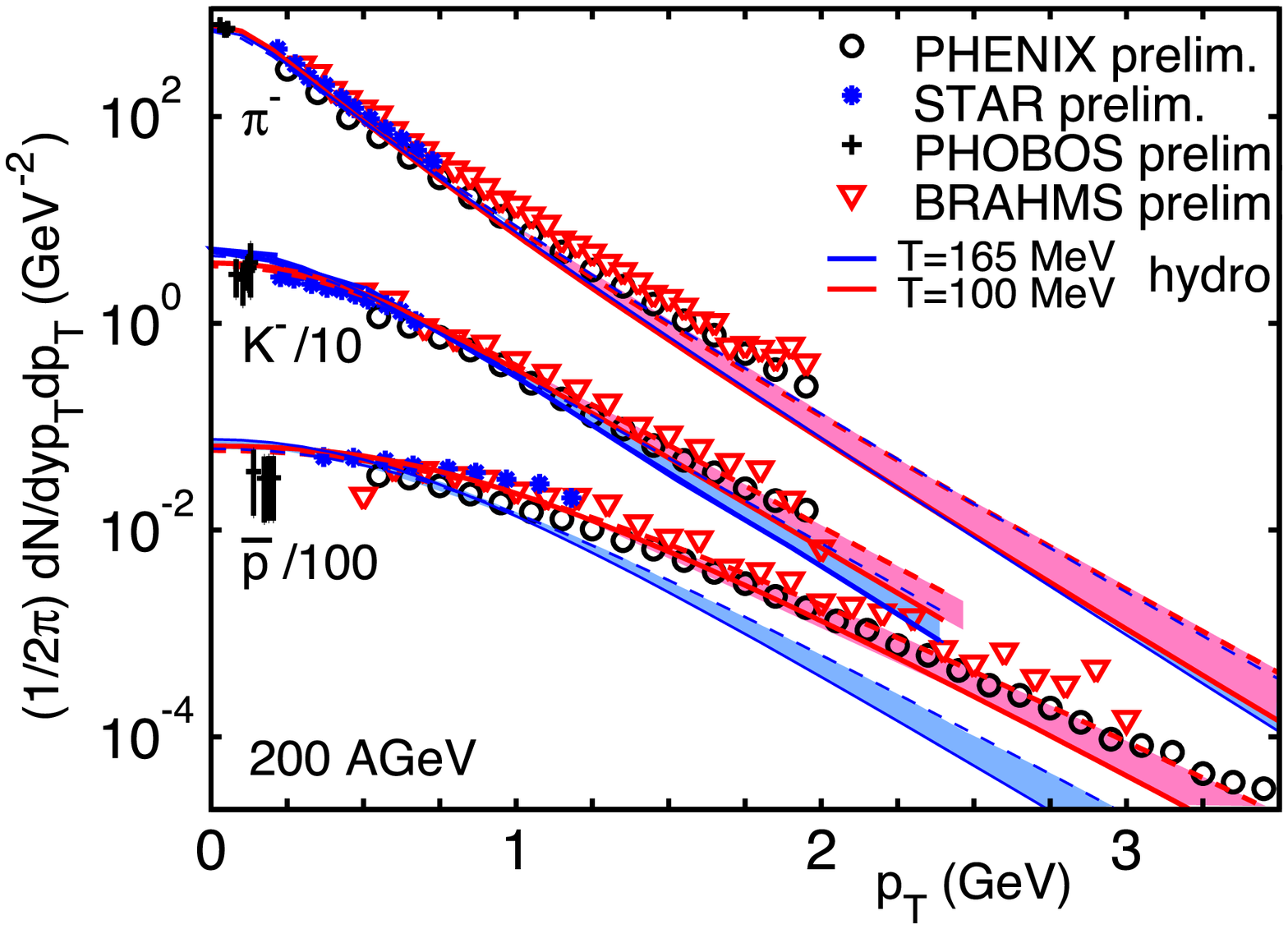}
\end{minipage}
\hspace*{3mm}
\begin{minipage}[t]{62mm}
\includegraphics[bb=20 32 513 405,width=62mm,height=44.8mm]{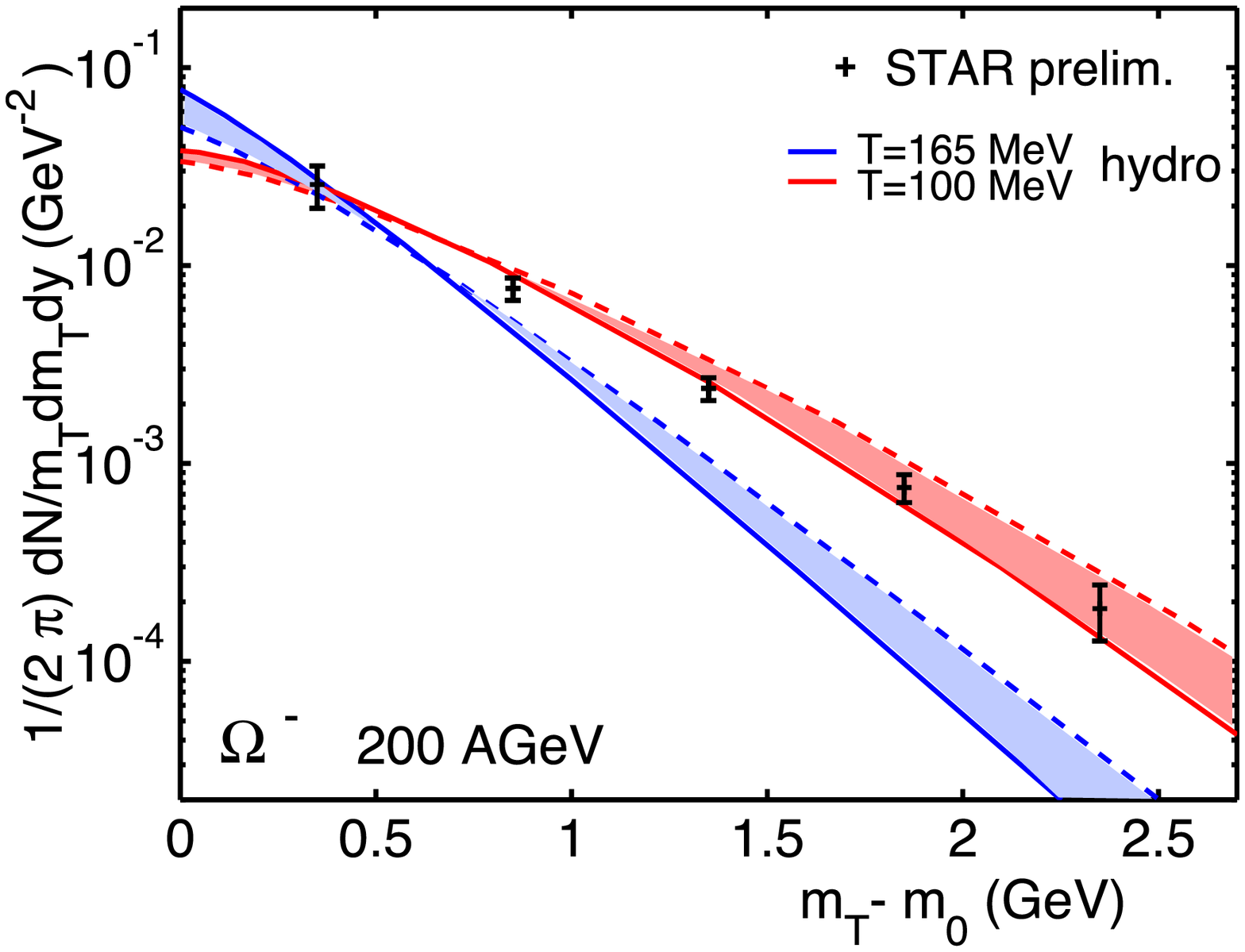}
\end{minipage}
\caption{\label{F1} 
(Color online) Negative pion, kaon, antiproton, and $\Omega$ spectra from 
central Au+Au collisions at $\sqrt{s}\eq200\,A$\,GeV, as measured by the
four RHIC experiments \cite{PHENIX03spec200,STAR03spec200,%
PHOBOS03spec200,BRAHMS03spec200,STAR03omega}. The curves show hydrodynamical 
calculations \cite{KR03,Kolb:2003dz} as described in the text.}
\end{figure}

For 130\,$A$\,GeV and 200\,$A$\,GeV Au+Au collisions at RHIC, these
predictions were surprisingly successful (see review in \cite{Kolb:2003dz}).
Figure \ref{F1} shows the fit to the (absolutely normalized) pion and 
antiproton spectra and,
%
%%%%%%%%%%%%%%%%%%%%%%%%% Fig. 1 %%%%%%%%%%%%%%%%%%%%%%%%%%%%%%%%%%%%%%%%%%%%
%%%%%%%%%%%%%%%%%%%%%%%%%%%%%%%%%%%%%%%%%%%%%%%%%%%%%%%%%%%%%%%%%%%%%%%%%%%%%%%
%
as examples, the predictions for the kaon and $\Omega$ spectra from 
200\,$A$\,GeV Au+Au collisions. Two sets of theoretical curves \cite{KR03}
are shown: the lower (blue) bands correspond to kinetic decoupling 
directly after hadronization at $T_{\rm cr}\eq165$\,MeV, whereas the 
upper (red) bands assume decoupling at $T_{\rm dec}\eq100$\,MeV. The 
width of the bands indicates the sensitivity of the calculated spectra 
to an initial transverse flow of the fireball already at the time of 
thermalization (for details see \cite{KR03}). The hydrodynamic model 
output shows \cite{KSH00} that it takes about 9-10\,fm/$c$ until the
fireball has become sufficiently dilute to completely convert to hadronic
matter, and another 7-8\,fm/$c$ to completely decouple. Figure~\ref{F1} 
shows clearly that by the time of hadronization hydrodynamics has not yet
generated enough radial flow to reproduce the measured $\bar p$ and $\Omega$
spectra; these heavy hadrons, which are particularly sensitive to radial
flow effects, require the additional collective ``push'' created by
resonant (quasi)elastic interactions during the fairly long-lived hadronic 
rescattering stage between $T_{\rm cr}$ and $T_{\rm dec}$. The late
hadronic stage of the collision thus plays an important role for the 
quantitative understanding of the shape of the single particle spectra
at RHIC and their radial flow. 

\begin{figure}[htb]
\begin{minipage}[t]{70mm}
\includegraphics*[bb=0 0 567 470,width=70mm,height=50mm]{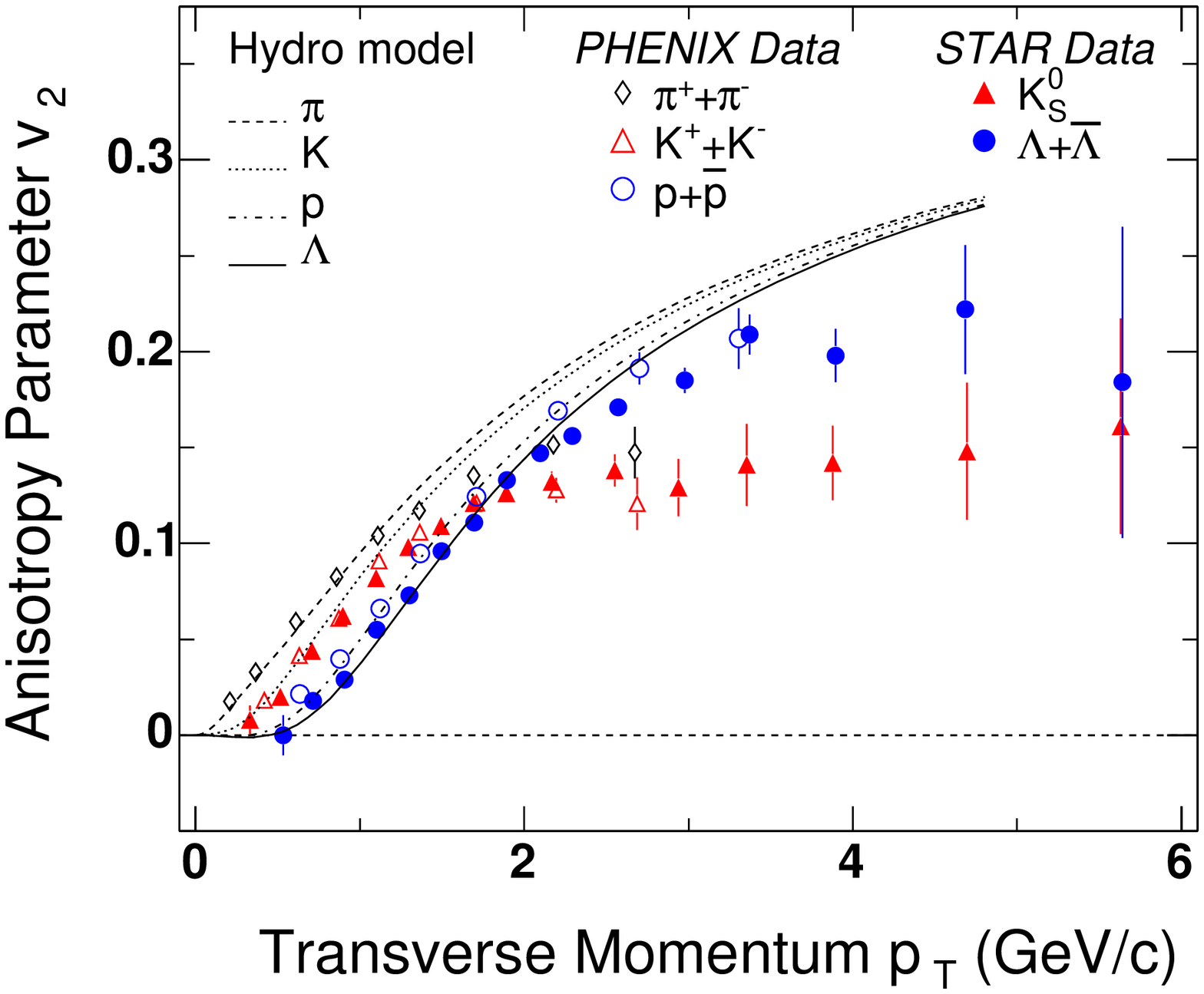}
\end{minipage}
\hspace*{-5mm}
\begin{minipage}[t]{62.5mm}
\includegraphics*[bb=61 198 568 594,width=62.5mm,height=47.35mm]{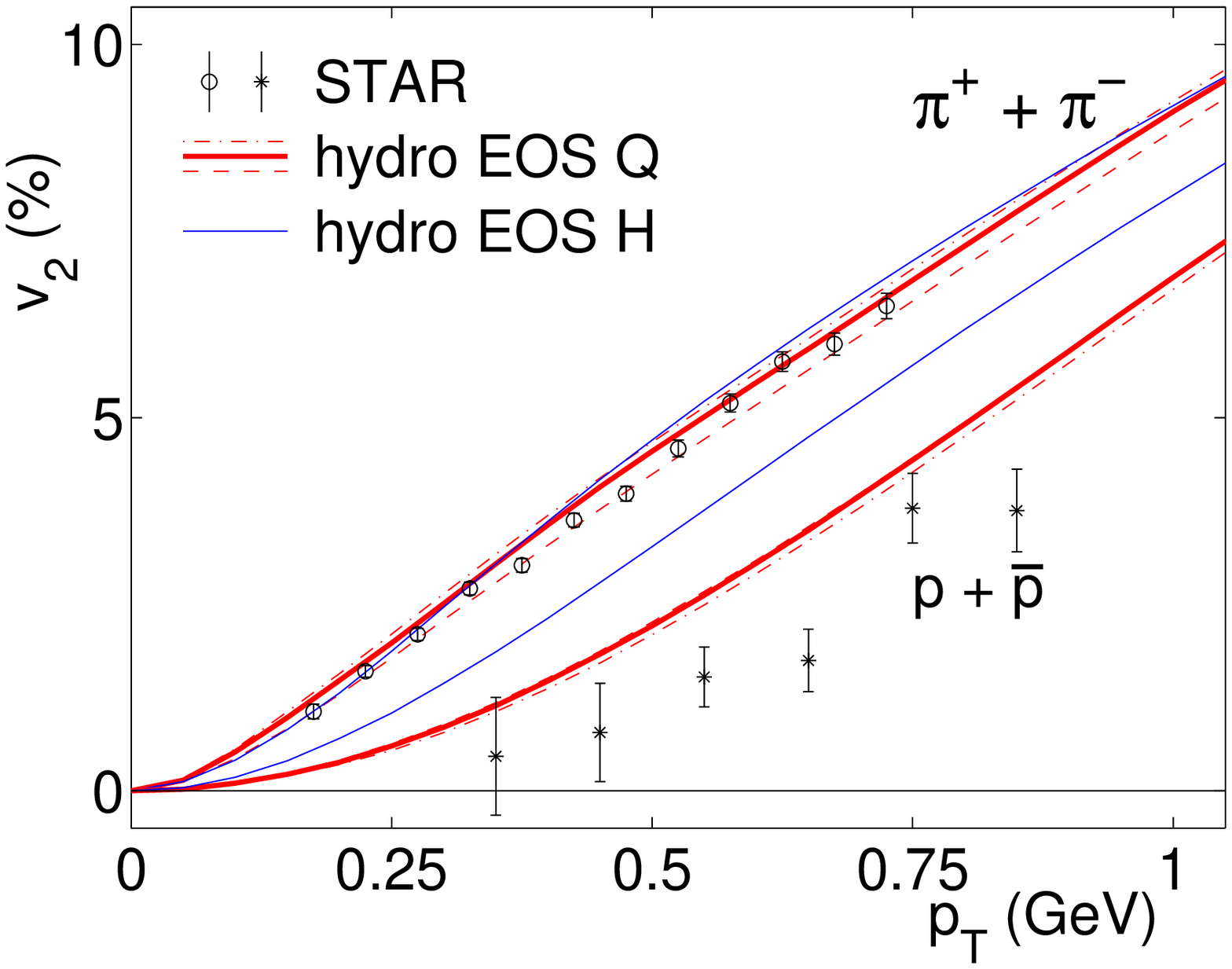}
\end{minipage}
\caption{\label{F2} 
(Color online) Differential elliptic flow $v_2(p_\perp)$ for several 
identified hadron 
species from minimum bias Au+Au collisions at 130 (right \cite{Adler:2001nb})
and 200\,$A$\,GeV (left \cite{Sorensen:2003kp}), compared with hydrodynamic 
predictions \cite{HKHRV01}. 
%\vspace*{-5mm}
}
\end{figure}

This is different for the elliptic flow which is driven by the spatial
anisotropy of the reaction zone in non-central collisions and the resulting
anisotropic pressure gradients: as shown in \cite{KSH00}, at RHIC energies,
for not too peripheral Au+Au collisions, the momentum anisotropy 
{\em saturates before the completion of hadronization}, due to the 
disappearance of the spatial deformation of 
the fireball. The measured $v_2$ thus reflects almost exclusively 
the early QGP dynamics and the phase transition region. Figure \ref{F2} 
shows that the corresponding hydrodynamic 
predictions work very well, at least for transverse momenta below about
1.5-2\,GeV/$c$ where most ($>99\%$) of the particles are emitted.
%
%%%%%%%%%%%%%%%%%%%%%%%%% Fig. 2 %%%%%%%%%%%%%%%%%%%%%%%%%%%%%%%%%%%%%%%%%%%%
%%%%%%%%%%%%%%%%%%%%%%%%%%%%%%%%%%%%%%%%%%%%%%%%%%%%%%%%%%%%%%%%%%%%%%%%%%%%%%%
%

Hydrodynamics predicts a characteristic rest mass dependence of $v_2$
\cite{HKHRV01} which is nicely reproduced by the data in Fig.~\ref{F2}. 
This ``fine structure'' constitutes com\-pel\-ling evidence that 
the hot matter produced at RHIC behaves like a thermalized fluid. The
``mass scaling'' of $v_2$ seen at low $p_\perp$ is quite distinct 
from the ``quark num\-ber scaling'' of $v_2$ seen at larger transverse 
momenta $p_\perp\gapp2-4$\,GeV/$c$ \cite{Sorensen:2003kp}; the latter is 
characteristic of a quark-coalescence mechanism for hadron production
which dominates in that $p_\perp$ range \cite{quarkcoal}. The mass 
scaling seen in the hydrodynamic regime is sensitive to the EOS of the
fireball matter: the right panel in Fig.~\ref{F2} shows two calculations,
one without a phase transition (EOS H), the other including the
quark-hadron transition (EOS Q). Although neither one fits the STAR 
proton data particularly well, the latter show a clear preference for 
the EOS including a phase transition.

%%%%%%%%%%%%%%%%%%%%%%%%%%%%%%%%%%%%%%%%%%%%%%%%%%%%%%%%%%%%%%%%%%%%%%%%%%%%
\section{Where ideal hydrodynamics breaks down}
%%%%%%%%%%%%%%%%%%%%%%%%%%%%%%%%%%%%%%%%%%%%%%%%%%%%%%%%%%%%%%%%%%%%%%%%%%%%

\begin{figure}[htb]
\begin{minipage}[t]{70mm}
\includegraphics*[width=70mm,height=52mm]{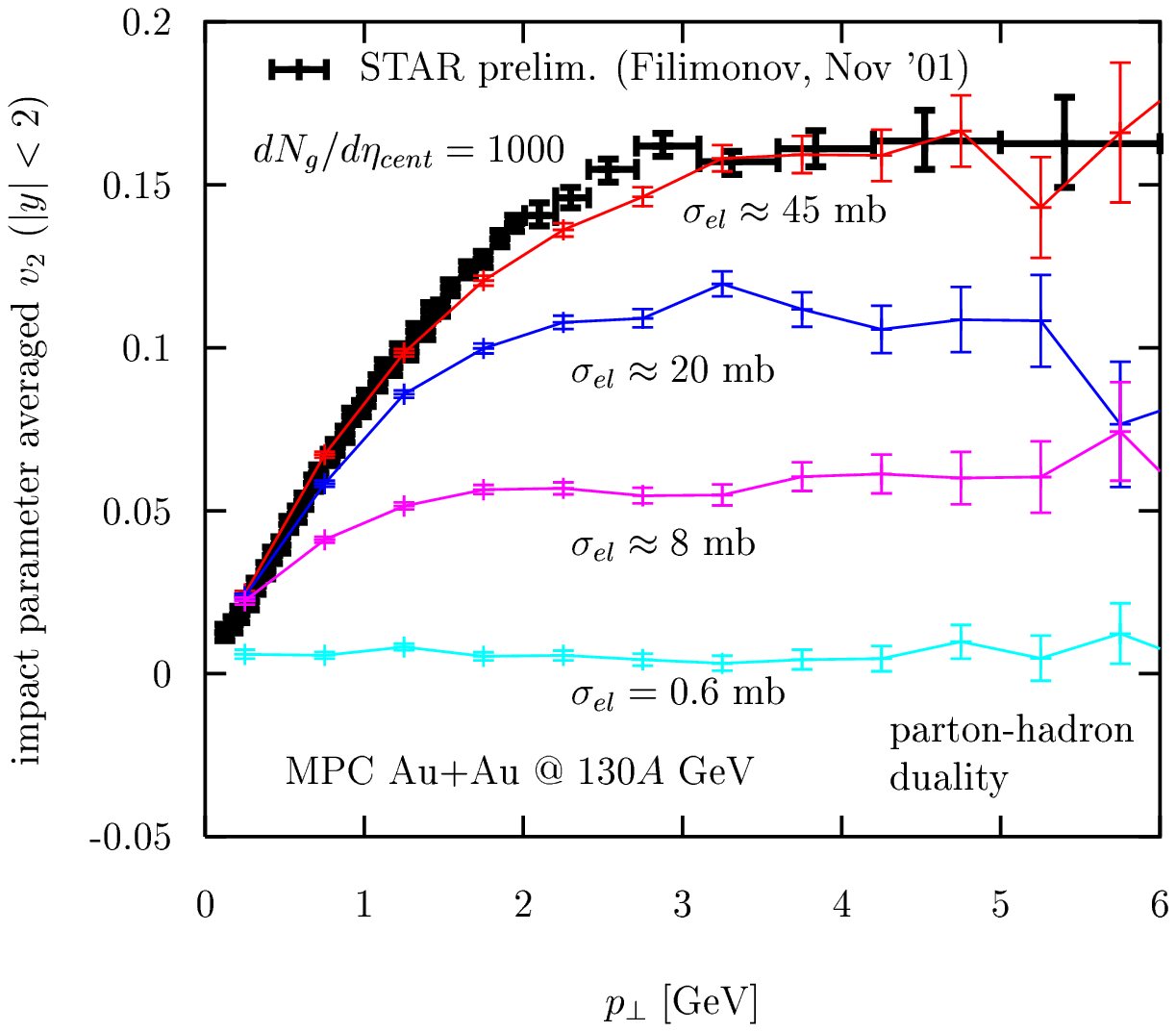}
\end{minipage}
\hspace*{-2mm}
\begin{minipage}[t]{62.5mm}
\includegraphics*[bb=0 -43 568 594,width=62.5mm,height=61mm,clip=]{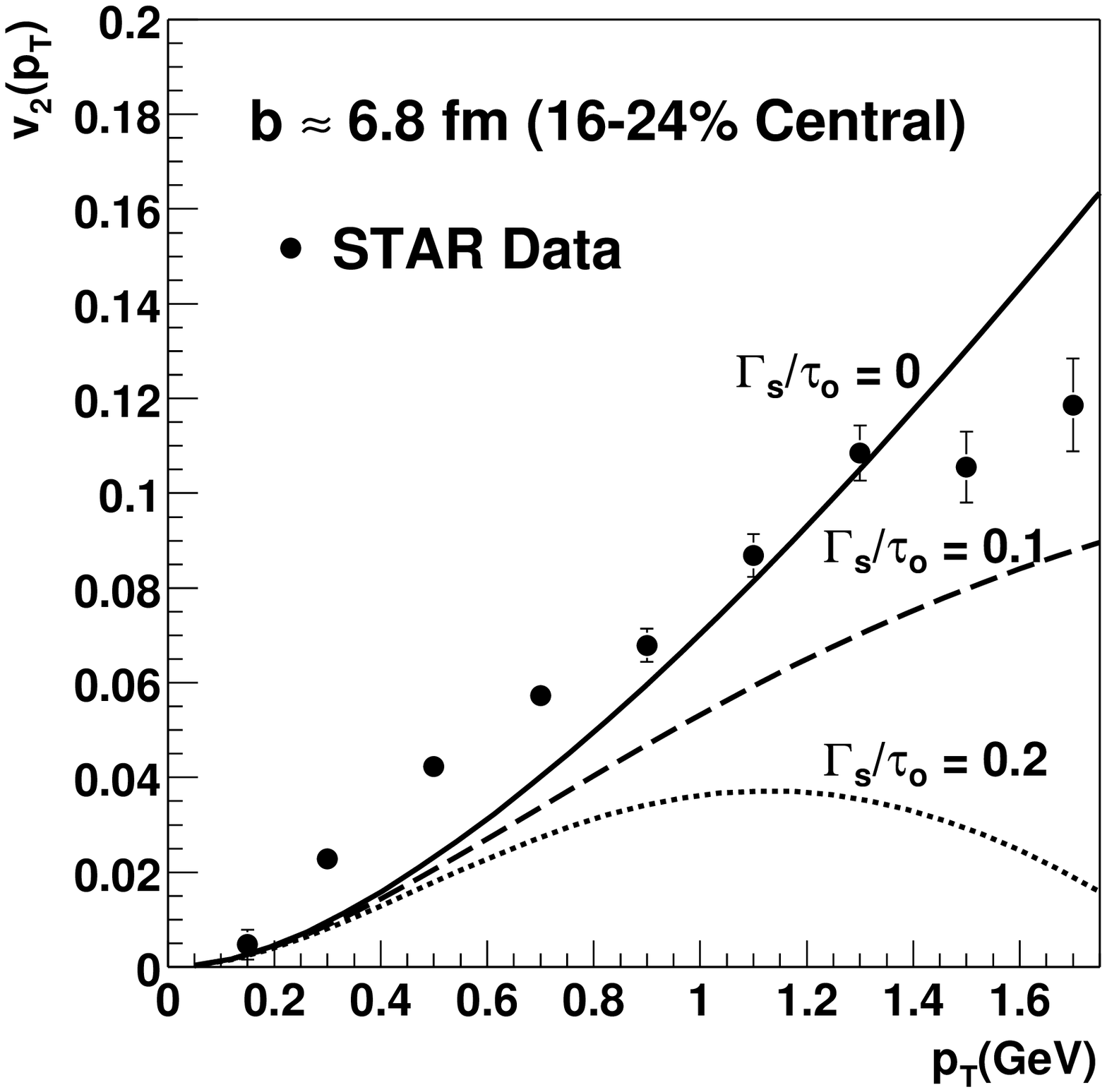}
\end{minipage}\\[-6mm]
\caption{\label{F3} 
(Color online) Left: Elliptic flow from a parton cascade 
\cite{Molnar:2001ux}, compared
with STAR data, for different parton-parton scattering cross sections.
Larger cross sections lead to smaller mean free paths. Right: Perturbative
effects of shear viscosity on the elliptic flow $v_2(p_\perp)$
\cite{T03} (see text for discussion).
%\vspace*{-5mm}
}
\end{figure}

The elliptic flow coefficient $v_2(p_\perp)$ is a measure for the 
(relatively small) differences between the transverse momentum spectra
with momenta pointing into and perpendicular to the reaction plane.
As such, it is more sensitive to deviations from ideal hydrodynamic 
behaviour than the overall shapes (slope parameters) of the transverse 
momentum spectra. Two model studies \cite{Heinz:2002rs,T03} showed
that $v_2$ reacts particularly strongly to shear viscosity.
As the mean free path of the plasma constituents (and thus the fluid's
viscosity) goes to zero, $v_2$ approaches the ideal hydrodynamic limit
from below \cite{ZGK99,Molnar:2001ux} (see Fig.\,\ref{F3}a). At higher 
transverse momenta it does so more slowly than at low $p_\perp$ 
\cite{Molnar:2001ux}, approaching a constant saturation value as seen
in the data rather than following the hydrodynamic almost linear rise as 
a function of $p_\perp$ shown by the lines in Fig.\,\ref{F2}. The 
increasing deviation from the%
%
%%%%%%%%%%%%%%%%%%%%%%%%% Fig. 3 %%%%%%%%%%%%%%%%%%%%%%%%%%%%%%%%%%%%%%%%%%%%
%\vspace*{-8mm}
%%%%%%%%%%%%%%%%%%%%%%%%%%%%%%%%%%%%%%%%%%%%%%%%%%%%%%%%%%%%%%%%%%%%%%%%%%%%%%%
%
ideal fluid limit for growing $p_\perp$ is qualitatively
consistent with the expected influence of shear viscosity: Teaney 
\cite{T03} showed that viscous corrections to the local thermal 
equilibrium distribution function increase in first order 
quadratically with $p_\perp$, leading to stronger and stronger
reduction of $v_2$ below the ideal fluid limit as $p_\perp$ grows
(see Fig.\,\ref{F3}b). From the results in Fig.\,\ref{F3}b Teaney
concluded that at RHIC the normalized sound attenuation length 
$\frac{\Gamma_s}{\tau} = \frac{4}{3T\tau}\frac{\eta}{s}$ (where $\eta$ is 
the shear viscosity, $T$ the temperature and $s$ the entropy density) 
cannot be much larger than about 0.1. This puts a stringent limit on 
the dimensionless ratio $\eta/s$, bringing it close to the recently
conjectured absolute lower limit for the viscosity of
$\eta/s\eq\hbar/(4\pi)$ \cite{son}. This would make the quark-gluon
plasma the most ideal fluid ever observed \cite{son}.

These arguments show that deviations from ideal fluid dynamics at
high $p_\perp$ must be expected, and that they can be large even
for fluids with very low viscosity. At which $p_\perp$ non-ideal 
effects begin to become visible in $v_2(p_\perp)$ can be taken 
as a measure for the fluid's viscosity. Figure \ref{F2}a, however, 
shows that mesons seem to deviate from the ideal fluid dynamical 
prediction earlier than baryons which follow the hydrodynamic 
curves to considerably larger $p_\perp$, thereby reaching 
significantly larger saturation values for $v_2$. Why should 
viscous effects be different for baryons and mesons? In fact, both
the $p_\perp$-value where $v_2$ begins to drop below the hydrodynamic 
prediction and the asymptotic saturation values of $v_2$ appear to be
independent of the hadron mass and only care about whether the hadron
is a meson or a baryon! 

This ``valence quark number scaling'' of $v_2$ at intermediate and high 
$p_\perp$ can be explained through quark coalescence \cite{quarkcoal}. 
This model predicts that the elliptic flow divided by the number
of valence quarks $n$, plotted against $p_\perp/n$, should be a 
universal function which reflects the partonic collective elliptic 
flow just before hadronization: 
$v_2^{\rm part}(p_\perp)\eq\frac{{v}_2^{\rm had}}{n}
\left(\frac{p_\perp^{\rm had}}{n}\right)$ \cite{quarkcoal}. For 
$p_\perp\gapp750$\,MeV, this scaling 
works extremely well \cite{Sorensen:2003kp}, at {\em all collision 
centralities} (note that $v_2$ is a strong function of centrality!).
Strange and non-strange partons seem to carry the same elliptic flow
\cite{Sorensen:2003kp}, consistent with strong rescattering and 
thermalization in the partonic phase. Below $p_\perp\sim750$\,MeV
the extracted partonic elliptic flow follows the hydrodynamically
predicted almost linear rise with $p_\perp$, and quark number scaling 
is violated and replaced by the hydrodynamically predicted mass scaling.
In summary, the critical value of $p_\perp$ where QGP viscosity manifests
itself in the partonic elliptic flow is ${\approx\,}750$\,MeV/$c$. What 
this implies for the value of the QGP shear viscosity $\eta$ remains to be 
seen, by comparing the data to simulations with a viscous relativistic 
hydrodynamic code.  

\begin{figure}[htb]
\begin{minipage}[t]{70mm}
\includegraphics*[bb=0 0 567 470,width=70mm,height=55mm]{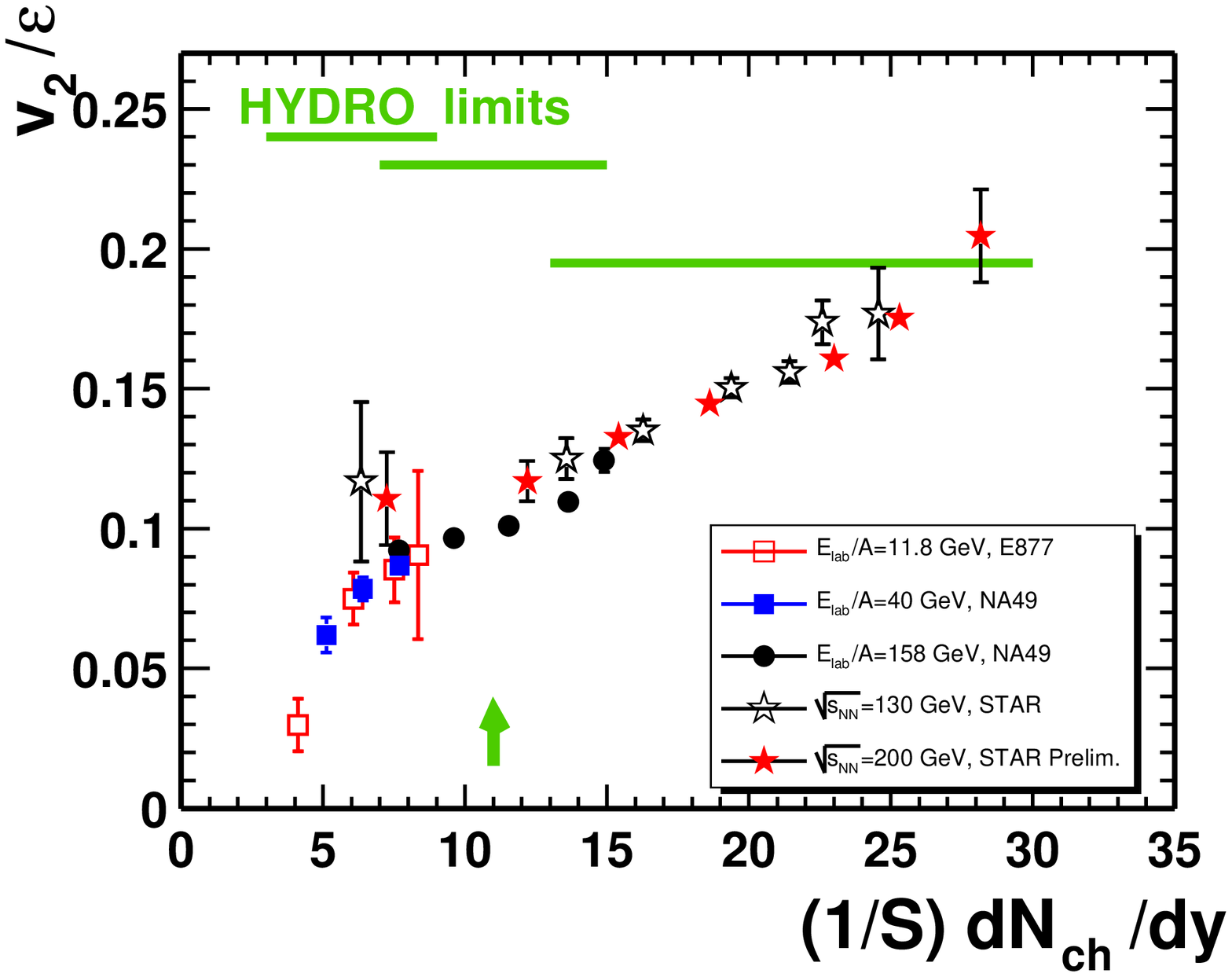}
\end{minipage}
\hspace*{-2mm}
\begin{minipage}[b]{62.5mm}
\includegraphics*[bb=0 -60 568 490,width=62.5mm,height=44.4mm]{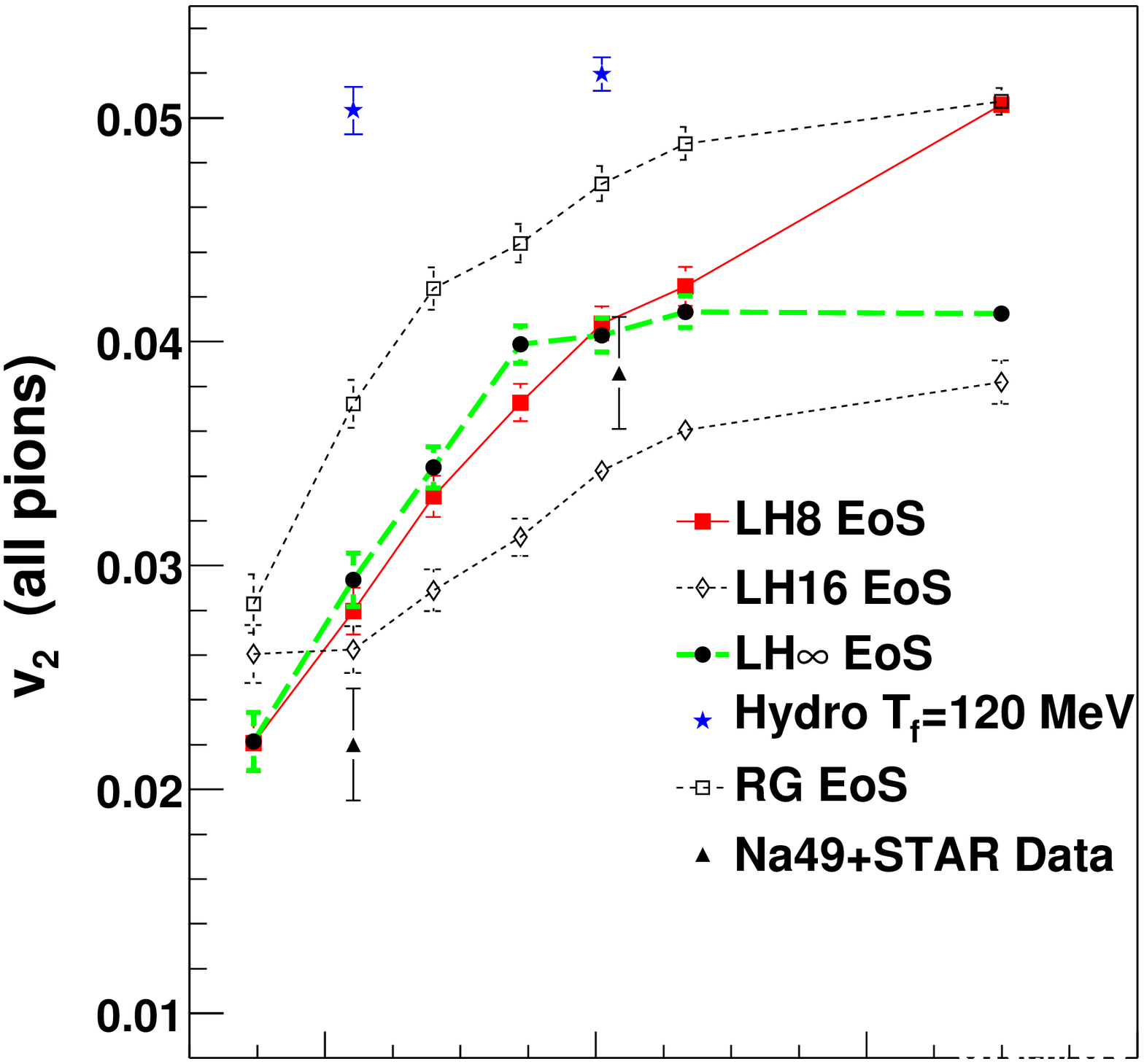}\\[-6mm]
\includegraphics*[bb=0 0 567 68,width=62.4mm,height=8mm]{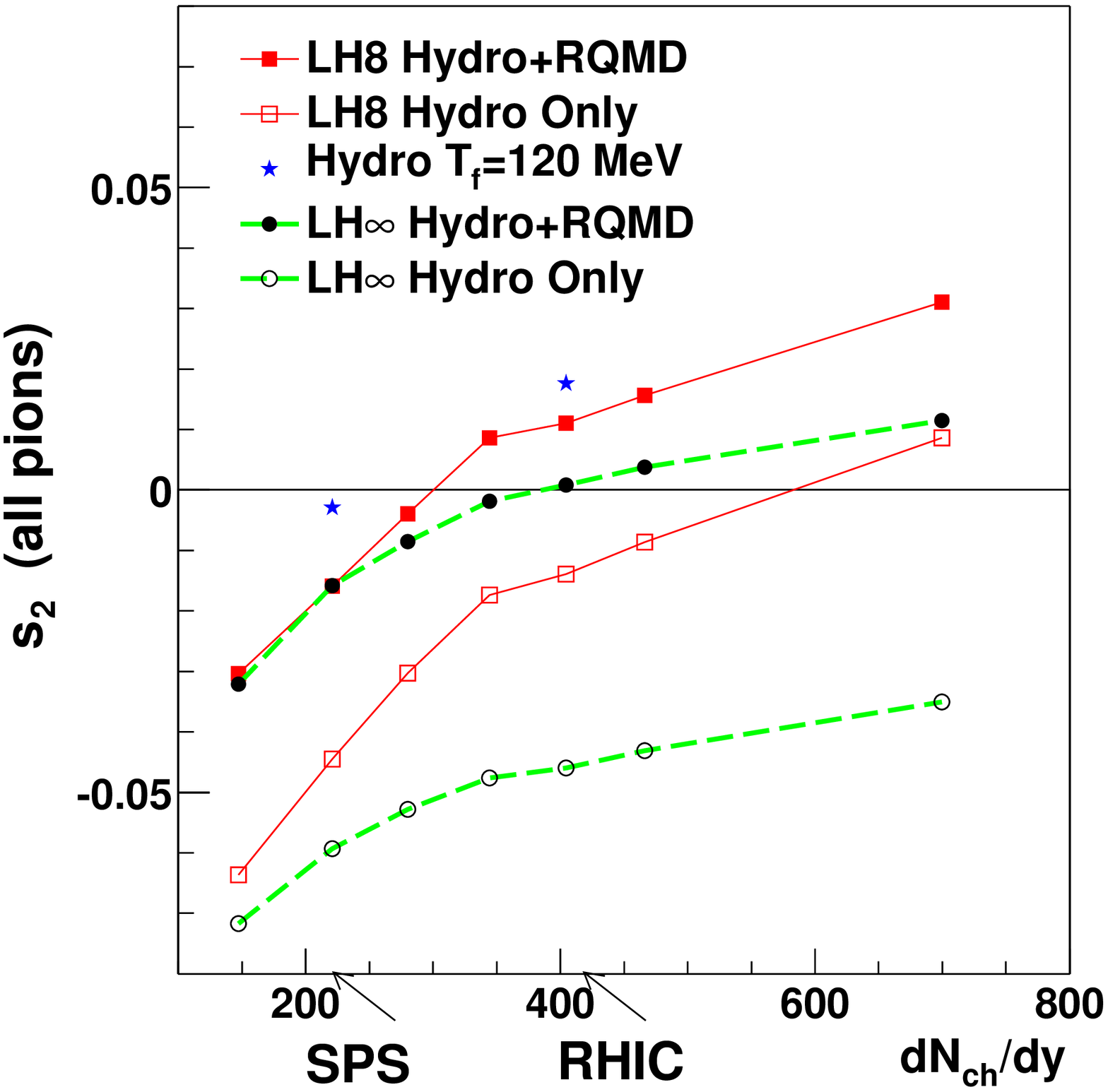}\\[-2.3mm]
\end{minipage}
\vspace*{-5mm}
\caption{\label{F4} 
(Color online) Left: Scaled elliptic flow $v_2/\epsilon$ (where $\epsilon$ 
denotes the initial spatial eccentricity) vs. the charged multiplicity 
density per
unit initial transverse overlap area $S$ \cite{NA49v2PRC}.
Right: Elliptic flow $v_2$ for minimum bias Au+Au collisions
at various collision energies, parametrized by the final charged
multiplicity density at midrapidity \cite{Teaney:2001av}. See text
for discussion.
}
\end{figure}

Hydrodynamics also fails to describe the elliptic flow $v_2$ in more 
peripheral Au+Au collisions at RHIC, in central and peripheral 
collisions at lower energies (for both see Fig.~\ref{F4}a), and in 
minimum bias collisions at RHIC at forward rapidities \cite{Hirano:2001eu}.
Whereas hydrodynamics predicts a non-monotonic beam energy dependence
of $v_2$ \cite{KSH00}, with largest values at upper AGS and lower
SPS energies (see Fig.~\ref{F4}a), somewhat lower values at RHIC and 
again larger values at the LHC, the data seem to increase monotonically
with $\sqrt{s}$.

%
%%%%%%%%%%%%%%%%%%%%%%%%% Fig. 4 %%%%%%%%%%%%%%%%%%%%%%%%%%%%%%%%%%%%%%%%%%%%
\vspace*{-5mm}
%%%%%%%%%%%%%%%%%%%%%%%%%%%%%%%%%%%%%%%%%%%%%%%%%%%%%%%%%%%%%%%%%%%%%%%%%%%%%%%
%

Such a monotonic rise is consistent with ``hybrid'' calculations by Teaney 
\cite{Teaney:2001av} (Fig.~\ref{F4}b) where the fireball undergoes ideal 
fluid dynamic evolution only to a point just below the hadronization 
phase transition, followed by hadronic {\em kinetic} evolution using 
the RQMD code until final decoupling. Figure~\ref{F4}b shows several 
curves corresponding to different equations of state during the 
hydrodynamic evolution (see \cite{Teaney:2001av} for details), with LH8 
%(an EOS with a first order phase transition with latent heat of 
% 0.8\,GeV/fm$^3$ separating the hadronic and QGP phases) 
being closest to the data. The difference 
between the points labelled LH8 and the hydrodynamic values at the
top of the figure is due to the different evolution during the 
late hadronic stage. Obviously, at lower collision 
energies and for impact parameters $b{\,\sim\,}7$\,fm (which dominate 
$v_2$ in minimum bias collisions), ideal fluid dynamics 
continues to build additional elliptic flow during the hadronic stage, 
in contrast to RQMD. The reason for the additional hadronic $v_2$    
in the hydrodynamic simulations is that at lower beam energies the 
initial energy densities are smaller and the fireball does not spend 
enough time
in the QGP phase for the spatial eccentricity $\epsilon$ to fully 
disappear before entering the hadron resonance gas phase. Anisotropic
pressure gradients thus still exist in the hadronic phase, and
hydrodynamics reacts to them according to the stiffness of the hadron
resonance gas EOS ($p\approx0.15e$). Teaney's calculations 
\cite{Teaney:2001av} show that RQMD reacts to these remaining 
anisotropies much 
more weakly, building very little if anny additional elliptic flow
during the hadronic phase. The hadron resonance gas, as modelled in
RQMD, is a highly viscous medium, unable to reach or maintain
local thermal equilibrium and therefore not behaving as an ideal fluid 
at all. The failure of the hydrodynamic model in peripheral collisions
at RHIC and in central and peripheral collisions at the SPS and AGS is
therefore {\em not} necessarily caused by the absence of an ideal
fluid QGP phase during the early expansion stage, but rather by
the highly viscous late hadronic stage which is unable to efficiently 
react to the remaining spatial fireball eccentricity. Similar arguments
hold at forward rapidities at RHIC 
\cite{Heinz:2004et} where the initial energy densities are also 
significantly smaller than at midrapidity while the initial spatial
eccentricities are similar.

The large hadron gas viscosity spoils one of the clearest experimental
signatures for the existence of the quark-hadron phase transition, the 
predicted non-monotonic beam energy dependence of $v_2$ \cite{KSH00}. 
The ideal fluid dynamic limit for $v_2$ scales with the speed of 
sound $c_s^2\eq\frac{\partial p}{\partial e}$. 
As one passes from the hadron resonance gas through the phase transition
into the QGP, $c_s^2$ varies from about 0.15 in the hadron gas to
about $\frac{1}{3}$ in the QGP, going through a minimum near zero in
the soft transition region. Ideal fluid dynamics maps this 
behaviour directly onto $v_2$ as one changes the initial energy
density by increasing the collision energy. As one comes down from
infinite beam energy, $v_2$ is therefore
predicted to first decrease (due to the softening of the EOS in the 
phase transition region) and then recover somewhat in the moderately 
stiff hadron gas phase. The hadron gas viscosity spoils this recovery,
leading to an apparently monotonous decrease of $v_2$ with falling
beam energy (Fig.~\ref{F4}). However, recent PHENIX data from Au+Au 
collisions at $\sqrt{s}\eq62\,A$\,GeV \cite{PHENIX62v2} indicate that 
this decrease may not be quite as monotonous as suggested by 
Fig.~\ref{F4}a. They show essentially constant elliptic flow over
the entire energy range explored at RHIC (from 62 to 200 $A$ GeV), 
decreasing only when going further down to SPS and AGS energies. While 
this is does not confirm the hydrodynamically predicted {\em rise} of 
$v_2$, it may still be a strongly diluted reflection of this predicted 
non-monotonic structure in the elliptic flow excitation function.
Obviously, many and more systematic hybrid calculations of the type 
pioneered by Teaney \cite{Teaney:2001av} are necessary to explore 
to what extent we can eventually prove the existence of a QCD phase 
transition using elliptic flow data.

We recently pointed out \cite{HK04} a very nice possibility to extend 
Fig.~\ref{F4}a towards the right, in order to confirm that for
higher initial energy densities the measured $v_2$ indeed settles 
on the ideal hydrodynamic curve. By colliding deformed uranium nuclei 
at RHIC, using the Zero Degree Calorimeters to select only full 
overlap collisions, and then exploiting the known binary 
collision component in the produced charged particle multiplicity
to select configurations with different initial spatial eccentricities
by cutting on multiplicity, one can extend the explored region in
Fig.~\ref{F4}a by about 60\% along the horizontal axis. In fact,
the lowest value for $\frac{1}{S}\frac{dN_{\rm ch}}{dy}$ obtainable
in this way is for U+U collisions in the side-on-side configuration,
which reproduce similar initial energy densities as central Au+Au 
collisions, but with a large initial eccentricity of 25\%. The opposite
limit is reached for U+U collisions in the nose-on-nose configuration
where the elliptic flow approaches zero but 
$\frac{1}{S}\frac{dN_{\rm ch}}{dy}$ is about 60\% higher than the 
largest value plotted in Fig.~\ref{F4}a.

%%%%%%%%%%%%%%%%%%%%%%%%%%%%%%%%%%%%%%%%%%%%%%%%%%%%%%%%%%%%%%%%%%%%%%%%%%%%
\section{Conclusions}
%%%%%%%%%%%%%%%%%%%%%%%%%%%%%%%%%%%%%%%%%%%%%%%%%%%%%%%%%%%%%%%%%%%%%%%%%%%%

The excellent reproduction of all aspects of single-particle hadron 
spectra measured in central and semicentral Au+Au collisions
at RHIC, including the elliptic flow and its fine structure, by 
relativistic hydrodynamics provides strong evidence for the creation
of a thermalized medium at high energy density $e{\,>\,}10$\,GeV/fm$^3$
with thermalization time $\tau_{\rm therm}{\,<\,}1$\,fm/$c$. The only
known thermalized state at such energy densities it the quark-gluon 
plasma. The observation of quark-number scaling of the elliptic flow
and other observables at intermediate $p_\perp$ indicates that deconfined
valence quarks play a dynamical role in hadron production, providing
at least indirect evidence for color deconfinement. The almost ideal
fluid behaviour of the expanding fireball shows that the QGP is a
strongly coupled plasma with fluid-like rather than gas-like properties
and with the lowest viscosity of any known liquid.
In contrast to the QGP, the hadron gas dominating the late expansion 
stage is highly viscous and does not behave like an ideal fluid. 
Hydrodynamics works for the first time quantitatively at RHIC
since the fireball dynamics is for the first time dominated by
the QGP.

We have reached an interesting stage where we are beginning to explore
the detailed properties of this new QGP state of matter. Clearly, 
many detailed studies, both theoretical and experimental, are 
required to achieve that goal. Systematic hydrodynamic studies,
combined with high statistics data on spectra and elliptic flow,
should help to further constrain the thermalization time
$\tau_{\rm therm}$ and equation of state. Hybrid hydro+cascade
calculations should be performed to isolate the non-ideal effects 
from the late hadronic stage. A 3+1-dimensional relativistic viscous 
hydrodynamic code must be developed to quantitatively constrain
the transport properties of the QGP (such as its viscosity) from 
the data, and thereby provide phenomenological constraints for 
theoretical efforts to calculate the latter from first principles.

%\smallskip

%\noindent{\bf Acknowledgement:} This work was supported by the 
%U.S. Department of Energy under Grant No. DE-FG02-01ER41190.
%\vspace*{-2mm}
%%%%%%%%%%%%%%%%%%%%%% References %%%%%%%%%%%%%%%%%%%%%%%%%%%%%%%%%%%%%%%%%%%
\section*{References}
%%%%%%%%%%%%%%%%%%%%%%%%%%%%%%%%%%%%%%%%%%%%%%%%%%%%%%%%%%%%%%%%%%%%%%%%%%%%%
{}
\end{document}